\documentstyle[12pt,openbib]{article}

\title
{ {\bf Lattice Discretization in  Quantum  Scattering }}

\author
{Sadhan K. Adhikari$^{1}$\thanks{John Simon Guggenheim Memorial Foundation 
Fellow}, 
T. Frederico$^{2}$, and R. M. Marinho$^{2}$\\
$^{1}$ Instituto de F\'\i sica Te\'orica, Universidade Estadual Paulista,\\
01405-900 S\~{a}o Paulo, S\~{a}o Paulo, Brasil \\
 $^{2}$  Instituto Tecnol\'ogico de Aeronautica,
        Centro T\'ecnico Aeroespacial,\\
            12228-900 S\~ao Jos\'e dos Campos, S\~ao Paulo, Brasil \\
}
\date{\today}
\begin{document}
\maketitle

\date{\today }

\begin{abstract}

The utility of lattice discretization technique is  demonstrated
for solving nonrelativistic quantum scattering problems and
specially for the treatment of ultraviolet divergences in these
problems with some potentials singular at the origin in two and
three  space dimensions.  This shows that lattice discretization
technique could be a useful tool for the  numerical soiution of
scattering problems in general.  The approach is illustrated in
the case of the Dirac delta function potential.

\end{abstract}
{PACS Numbers 03.80.+r, 11.15.Ha, 11.10.Gh  }

The technique of discretization on lattice  (hereafter called
lattice technique) \cite{lat,lat1,lat2} has been successfully
used to deal with ultraviolet divergences in gauge field
theoretic problems in perturbative expansion, specially those in
quantum electrodynamics (QED) and quantum chromodynamics (QCD).
These ultraviolet divergences in perturbative quantum field
theory can be eliminated by lattice technique  to yield a scale.
Except in some simple cases the lattice-regularized perturbative
series can not be summed up and this makes it difficult to draw
conclusions about the full solution. The renormalization group
(RG) equations \cite{wil,ryd}, on the other hand, yield many
general properties of the full solution from the
lattice-regularized results of the perturbative expansion.

The lattice technique represents a mathematical trick, which
removes the ultraviolet divergences by introducing a cutoff in a
regularized  Green function. As with any regulator, it is
removed after renormalization. The physical observables are then
obtained in the continuum limit, where the lattice spacing is
taken to be zero.

Ultraviolet divergences also appear in the nonrelativistic quantum scattering
problems for potentials with certain singular behavior at short distances
\cite{wei,ad,tar1,beg} in two and three  dimensions.  In one dimension these
divergences are absent.  We show that the application of the lattice
technique to these potential models  leads to a scale and finite physical
observables after the continuum limit is taken by the usual renormalization
procedure. The present work is written in a pedagogic style so that it
clarifies all the subtleties of lattice technique in a simple nonrelativistic
problem and should serve as an introduction to the study of lattice technique
in a complicated field theoretic problem.

Recently, there have been discussions on  renormalization in configuration
\cite{tar1} and momentum \cite{wei,ad,beg} spaces for potential
scattering with the Dirac delta, contact, or zero-range potential.  In this
work the  lattice technique is used  for potential scattering with delta
potential in two and three dimensions.  In both cases there are ultraviolet
divergences.  In the two dimensional case the divergence is logarithmic in
nature, whereas in the three-dimensional case it is linear. The
lattice-regularized result is finally renormalized and the RG equation
written.

The  present potential scattering problem permits analytic solution and is
infinitely simpler than gauge field theories of QED and QCD where the lattice
technique is usually applied.  Hence the present study will allow us to
understand the subtleties of this approach.  In gauge field theories
particles can be spontaneously created and destroyed and the discretization
is done in the four-dimensional Euclidean space.  In  potential scattering
particle number is conserved and we directly discretize the  time-independent
Schr\"odinger equation for relative motion in three-dimensional Euclidean
space.

The present analytic investigation
with delta potential shows the subtleties of the lattice technique  
and demonstrates that this approach
can be used for a numerical solution of nonrelativistic quantum 
scattering problems in general, not only for two particles but for 
several particles. In this analytic study   we calculate the nonrelativistic
Green function  and the $t$ matrix on the lattice. 
The numerical study remains one to be attempted in the future. 

There is another interest to study the nonrelativistic  scattering with delta
potential in two dimensions. This problem  can be considered to be a good
model of the ultraviolet structure  and high energy behavior of
$\lambda\phi^4$ field theory \cite{ryd,tar1,beg}. Both problems have
ultraviolet logarithmic divergences, require regularization, are
perturbatively renormalizable, collapse for attractive interaction but are
asymptotically free, etc.

We discuss $S$-wave potential scattering with the delta potential.  The
partial-wave Lippmann-Schwinger equation for the scattering amplitude
$T(p,q,k^2)$ in ${\cal D}$  dimensions at c.m.  energy $k^2$ is given by
\begin{eqnarray}
T(p',p,k^2) = V(p',p)+\int d^{\cal D}q V(p',q) 
G(q;k^2)T(q,p,k^2),
\label{2}
 \end{eqnarray} with the free Green function $G(q;k^2)=(k^2-q^2+i0)^{-1},$ in
units $\hbar=2m=1$, where $m$ is the reduced mass. The integral in Eq.
(\ref{2}) is  over the relevant $S$-wave phase space, e.g., we take
$d^3q\equiv 4\pi q^2dq$ and $d^2q\equiv 2\pi qdq$ with $q$ varying from 0 to
$\infty$. For the delta potential $V(p',p) = \lambda$, and
\begin{equation}
T(p',p,k^2)=[\lambda^{-1}-I(k)]^{-1},
\label{4}
\end{equation}
with $I(k)= \int d^{\cal D}qG(q;k^2).$ The integral $I(k)$ possesses
ultraviolet divergence for ${\cal D} > 1$. For  ${\cal D}=3$ (2) this
divergence is linear (logarithmic) in nature.  Finite result for the $t$
matrix of Eq.  (\ref{4}) can be obtained only if  $\lambda^{-1}$ also
diverges in a similar fashion and cancels the divergence of $I(k)$.

The solution of the problem can be achieved by discretizing the full
Schr\"odinger equation on lattice and finding its solution numerically.
Instead, as this problem permits analytic solution, we discretize the free
Schr\"odinger equation on lattice and evaluate the lattice-regularized free
Green function. With the lattice-regularized Green function the ultraviolet
divergences are avoided. In contrast to the lattice discretization of gauge
field theories, where one works in terms of Lagrangian densities and path
integrals \cite{lat1,lat2}, in the present problem it is convenient to work
in terms of the following time-independent Schr\"odinger equation for
relative motion
\begin{equation}
\nabla^2_{\bf r} \phi({\bf r}) + k^2 \phi({\bf  r}) =0,\label{10}
\end{equation}
where the space vector $ {\bf r} \equiv (x_j), j=1,...,{\cal D}.$ The present
mathematical treatment  is much simpler than, but similar to,  that in field
theory \cite{lat1}.  For our purpose we consider the ${\cal D}$-dimensional
lattice of spacing $a$.  The transition from the continuum to the discrete
lattice  is then effected  by making the following substitutions
\cite{lat,lat1}
\begin{eqnarray}
x_j & \to & {\bf n}a \equiv n_j a, j= 1,..., {\cal D},\nonumber   \\
\phi({\bf r}) & \to & \phi_{\bf n} \equiv \phi({\bf n} a)\nonumber \\
\nabla^2_{\bf r} \phi({\bf r}) & \to & 
a^{-2}\sum_{j=1}^{\cal D} \biggr[\phi({\bf n} a+ {\bf \hat j} a) +\phi({\bf
n} a- {\bf \hat j} a) -  2\phi({\bf n} a)\biggr].\nonumber
\end{eqnarray}
Here the space coordinate  is discretized by ${\bf r }= {\bf  n} a$ and $\bf
\hat j$ is the unit vector in direction $j$.  The individual component $n_j$
assumes only a finite number $N$ of independent values. Outside this range
the lattice is assumed to be periodic, so that the $n$th site can be
identified with the $(n+N)$th site. The active part of the lattice has
$N^{\cal D}$ sites.

After discretization,  the Schr\"odinger equation (\ref{10}) becomes the
matrix equation
\begin{eqnarray}
\sum_{\bf m} K_{\bf n \bf m }\phi _ {\bf m}  = 0,\label{11}
\end{eqnarray}
where
\begin{eqnarray}
K_{\bf n \bf m }  =  a^{-2} \sum_{j=1}^{\cal D} \biggr[\delta_{\bf n + \hat
j, \bf m} +\delta_{\bf n - \hat j, \bf m} +   (a^2k^2_j- 2)
\delta_{\bf n,\bf m}\biggr].
\end{eqnarray}
Comparing Eqs. (\ref{10}) and (\ref{11}) we realize that $K_{\bf n \bf m }$
is the discretized version of the operator $(\nabla^2_{\bf r}+k^2)$.  Hence,
the free Green function is the inverse of this operator, defined by
\begin{equation}
\sum_{\bf m}  K_{\bf n \bf m } (K^{-1})_{\bf m \bf l } = \delta_{\bf n \bf l }.
\end{equation}
This inverse operator can be evaluated analytically by working in momentum
space where the ${\cal D}$-dimensional Kr\"oneker $\delta$ functions are
represented as
\begin{equation}
\delta_{\bf n \bf m}= \biggr(\prod_{l=1}^{\cal
D}\int_{-\pi}^{\pi}\frac{d\tilde  q_l}{(2\pi)^3}\biggr)e^{i\bf \tilde q \cdot
(\bf n-\bf m)},\label{41}
\end{equation}
where $\bf \tilde  q$ is a dimensionless wave number defined by ${\bf \tilde
q} = a\bf q$ with components $\tilde q_j$.  The integration is restricted to
the first Brillouin zone $-\pi\le \tilde  q_l \le \pi$.  In the continuum
limit one has the following relations for the phase spaces
\begin{eqnarray}
\int d^{\cal D}q & \equiv &
\lim_{a\to 0}\biggr( \prod_l^{\cal D}\frac{1}{a^{\cal D}}
\int_{-\pi}^{\pi}\frac{d\tilde q_l }{(2\pi)^3} \biggr)\nonumber \\
& = & \lim_{a\to 0}\biggr( \prod_l^{\cal D}
\int_{-\pi/a}^{\pi/a}\frac{d q_l}{(2\pi)^3}  \biggr).\label{42}
\end{eqnarray}

Using the Fourier representation (\ref{41}), the matrix $K$ can be written as
\begin{eqnarray}
K_{\bf n \bf m } & = &a^{-2}\biggr(\prod_{l=1}^{\cal
D}\int_{-\pi}^{\pi}\frac{d\tilde  q_l}{(2\pi)^3}\biggr)e^{i\bf
\tilde  q \cdot  (\bf n-\bf m)}\nonumber 
\sum_{j=1}^{\cal D}\biggr[ e^{i\bf \tilde  q \cdot \hat j}+
e^{-i\bf \tilde  q
\cdot \hat j} + (a^2 k_j ^2 - 2)\biggr],\nonumber \\ & = &
a^{-2}\biggr(\prod_{l=1}^{\cal D}\int_{-\pi}^{\pi}\frac{d\tilde
q_l}{(2\pi)^3}\biggr)e^{i\bf
\tilde  q \cdot  (\bf n-\bf m)}
\biggr[(a^2 k^2- 2{\cal D}) +\sum_{j=1}^{\cal D} 2
\cos \tilde  q_j\biggr].
\label{51}\end{eqnarray}
 The  inverse of the matrix  $K$ is now determined by
\begin{eqnarray}
(K^{-1})_{\bf n \bf m }  = a^{2}
\biggr(\prod_{l=1}^{\cal
D}\int_{-\pi}^{\pi}\frac{d\tilde  q_l}{(2\pi)^3}\biggr)e^{i\bf
\tilde  q \cdot  (\bf n-\bf m)} 
\biggr[(a^2 k^2- 2{\cal D}) +\sum_{j=1}^{\cal D} 2
\cos \tilde  q_j\biggr]^{-1}.
\end{eqnarray}
This result leads to  the following lattice-regularized out\-going-wave Green
function
\begin{eqnarray}
G_R(  q,a;k^2) =  a^{2}\biggr[(a^2 k^2- 2{\cal D}) +
\sum_{j=1}^{\cal D} 2 \cos \tilde  q_j+i0\biggr]^{-1}.\label{rg}
 \end{eqnarray} With this Green function  there is no ultraviolet divergence
for $a \ne 0$. The imaginary part of this Green function guarantees unitarity
for outgoing-wave scattering.  In the limit $a \to 0 $, the regularized Green
function reduces to the free Green function: $\lim_{a\to 0 }G_R(q,a;k^2)=
G(q;k^2)$.

Using the lattice-regularized Green function  (\ref{rg}), the $t$ matrix
(\ref{4}) can  be rewritten as
\begin{eqnarray}
T(k,\lambda(a),a)& = &[\lambda^{-1}(a)-I_{R}(k,a)]^{-1} ,
\label{8} \end{eqnarray}
  where \begin{eqnarray} I_R(k,a)& \equiv & \int d^{\cal D}q
G_R(q,a;k^2)\nonumber \\ & = & a^{(2-{\cal D} )}\biggr(\prod_{l=1}^{\cal
D}\int_{-\pi}^{\pi}\frac{d\tilde q_l}{(2\pi)^3}\biggr)
\biggr[(a^2 k^2- 2{\cal D}) +\sum_{j=1}^{\cal D} 2
\cos \tilde q_j+i0\biggr]^{-1},
\end{eqnarray}
is a convergent integral for a finite lattice spacing $a$.  In Eq. (\ref{8})
the redundant momentum labels $p,p'$ of the $t$ matrix have been suppressed,
and the explicit dependences of the $t$ matrix on  $a$ and $\lambda(a)$ have
been introduced.  As $a \to 0$, however, this integral develops the original
ultraviolet divergence.  Explicitly,
\begin{eqnarray}
\lim_{a \to 0} I_R(k,a) & = & -[c/a+2\pi^2ik], {\cal D} =3, \\
\lim_{a \to 0} I_R(k,a) & = & 2 \pi\ln(ak)-i\pi^2, {\cal D} =2,
\end{eqnarray} 
where 
\begin{equation}
c\equiv \biggr(\prod_{l=1}^{\cal
D}\int_{-\pi}^{\pi}\frac{d\tilde q_l}{(2\pi)^3}\biggr) \biggr[6- \sum_{j=1}^3
2 \cos \tilde q_j\biggr]^{-1},
\nonumber  
\end{equation}
is a real finite definite integral.

   Finite results for physical magnitudes, as $a \to 0$, are obtained from
Eq. (\ref{8})  if the coupling $\lambda$ is also replaced by the so
called bare coupling $\lambda(a)$ as in this equation. The bare coupling can,
for example, be  defined by
\begin{eqnarray}
\lambda^{-1}(a) & = &  - [c/a+2\pi^2\Lambda_0], {\cal D } =3,\label{x} \\
 &  = & 2\pi [\ln (a\Lambda_0)], {\cal D} = 2, \label{y}
\end{eqnarray}
where $\Lambda_0$   is the physical scale of the problem and characterizes
the strength of the interaction. The quantities $\lambda^{-1}(a)$ of Eqs.
(\ref{x}) and (\ref{y}) have the appropriate divergent behavior, as $a
\to 0$, and cancel the divergent part of $I_R(k,a)$ in Eq. (\ref{8}).

Regarding the lattice merely as an ultraviolet regulator or cutoff (lattice
spacing $a$), finally, we must take the continuum limit: $a\to 0$.  For the
present delta potential this limit can be taken analytically. In a general
problem the limit has to be taken numerically. Both ways are illustrated
below. After this limit is taken the observables should approach their
physical values.  The question of renormalization is intimately related to
the removal of the regulator and prediction of physical observables.

Next the  $a \to  0$ limit is taken analytically  in Eq. (\ref{8}) and then
we turn to the question of renormalization.  With the present  regularization
procedure one has for the lattice-renormalized $t$ matrix
\begin{eqnarray}
T_R(k,\lambda_R({\cal A}),{\cal A})& = &[\lambda^{-1}_R({\cal A})-{\cal
I}_{R}(k,{\cal A})]^{-1} ,
\label{200} \end{eqnarray} with
\begin{eqnarray}
{\cal I}_{R}(k,{\cal A})& = &\lim_{a\to 0}[{ I}_{R}(k,a)-I_R(i/{\cal
A},a)],\label{202}\\ \lambda_R^{-1}({\cal A}) &   = & \lim_{a \to 0}
[\lambda^{-1}(a) -I_R(i/{\cal A},a)],
\label{207} \end{eqnarray}
where ${\cal A}$  is  the lattice-renormalization scale of the problem.  This
scale ${\cal A}$ should be contrasted with the physical scale $\Lambda_0$ of
Eqs. (\ref{x}) and (\ref{y}).  In Eq. (\ref{200})  the explicit dependence of
the $t$ matrix on both ${\cal A}$ and the lattice-renormalized coupling
$\lambda_R({\cal A})$ has been exhibited.

After taking the $a\to 0$ limit in Eq. (\ref{202}) and using Eq. (\ref{42}), 
the following lattice-renormalized function is obtained
\begin{eqnarray}
{\cal I}_{R}(k,{\cal A})& = & \lim_{a \to 0}
a^{(2-{\cal D} )}\biggr(\prod_{l=1}^{\cal
D}\int_{-\pi}^{\pi}\frac{d\tilde q_l}{(2\pi)^3}\biggr)\nonumber\\
&\times&
\frac {-a^2({\cal A}^{-2}+k^2)}
{ [\{a^2 k^2- 2{\cal D} +\sum_{j=1}^{\cal D} 2
\cos \tilde q_j+i0][-a^2 /{\cal A}
^2- 2{\cal D} +\sum_{j=1}^{\cal D} 2
\cos \tilde q_j] } \nonumber \\
& = & \int d^{\cal D} q \frac {{\cal A}^{-2}+k^2} {(k^2-q^2+i0)({\cal
A}^{-2}+q^2)}.
\end{eqnarray}
Consequently,
\begin{eqnarray}
{\cal I}_{R}(k,{\cal A})& = & -2\pi^2(ik+1/{\cal A}), {\cal D }=3,\\ {\cal
I}_{R}(k,{\cal A})& = & 2\pi\ln(k{\cal A})-i\pi^2, {\cal D }=2,
\end{eqnarray}
In Eq. (\ref{207}), if integrals $I_R$ are evaluated and the trivial limit $a
\to 0$ taken, we get
\begin{eqnarray}
\lambda_R({\cal A}) & = &  -[2\pi^2 /{\cal A}+2\pi^2\Lambda_0]^{-1}, {\cal
D } =3,\label{x1} \\ &  = & [2\pi \ln ({\cal A}\Lambda_0)]^{-1}, {\cal D} =
2. \label{y1}
\end{eqnarray}
The lattice-renormalized coupling for two scales ${\cal A}$ and ${\cal A}_0$
are related by the flow equations:
\begin{eqnarray}
\lambda_R^{-1}({\cal A})+2\pi^2/{\cal A} & = & \lambda_R^{-1}({\cal
A}_0)+2\pi^2/{\cal A}_0, \label{x2} \\ \lambda_R^{-1}({\cal A})-2\pi\ln{\cal
A} & = &
\lambda_R^{-1}({\cal A}_0)-2\pi\ln{\cal A}_0,\label{y2}
\end{eqnarray}
for ${\cal D} =3$ and 2, respectively.  The flow equations are independent of
the renormalization scheme.

The present scattering model permits analytic solution and for  ${\cal D} =
3$ and 2 the exact lattice-renormalized $t$ matrices of Eq.  (\ref{200}) are
given, respectively, by
\begin{eqnarray}
T_R(k,\lambda_R({\cal A}),{\cal A}) & = &  [\lambda_R^{-1}({\cal
A})+2\pi^2(1/{\cal A}+ik)]^{-1}, \\ & = & [\lambda_R^{-1}({\cal
A})-2\pi\ln(k{\cal A})+i\pi^2]^{-1}.
\end{eqnarray}   
Explicitly, using  definitions (\ref{x1}) and (\ref{y1}) for the renormalized
coupling, these lattice-renormalized $t$ matrices can be written as
\begin{eqnarray}
T_R(k,\lambda_R({\cal A}),{\cal A}) & = & [2\pi^2(ik-\Lambda_0)]^{-1}, {\cal
D } =3\\
\label{A}  & = &  
 [-2\pi\ln(k/\Lambda_0)+i\pi^2]^{-1}, {\cal D }=2.
\end{eqnarray}
These $t$ matrices depend on $\lambda_R({\cal A})$, but not on ${\cal A}$,
that is the explicit and implicit (through $\lambda_R({\cal A})$) dependences
of the $t$ matrix on ${\cal A}$ cancel. Physics is determined by the value of
$\lambda_R({\cal A})$ at an arbitrary value of ${\cal A}$ \cite{tar1}.  For
${\cal D}= $3, the physical scale $\Lambda_0$ is related to the scattering
length $a_0$ by $a_0= -1/\Lambda_0$.  For ${\cal D}=$ 2, $\Lambda_0$ can also
be related to the scattering length \cite{2d}.

Next we write the RG equation for this problem and show how the limit $a \to
0$ can be taken  numerically.  In this limit the lattice-renormalized $t$
matrix is independent of $a$, so is invariant under the group of
transformations ${ a}  \to {\exp}(s)a$, which form the RG.  It is convenient
to work in terms of the dimensionless coupling, $g(a)$, defined by
\begin{eqnarray}
g(a) & \equiv &  c\lambda(a)/a, {\cal D}=3, \label{355}  \\ &  \equiv &
2\pi\lambda(a), {\cal D}=2. \label{356} \end{eqnarray} The renormalization
condition in the $a\to 0$ limit can be expressed as \cite{lat}
\begin{equation} 
a \frac{d}{da} T(k,g(a),a) =0,
\label{360}\end{equation}
or,
\begin{equation}
\biggr[ a \frac{\partial}{\partial a}+ \beta(g)\frac{\partial}
{\partial g}
\biggr]T(k,g(a),a)=0,
\label{370}\end{equation}
where the RG function $\beta(g)$ is defined by
\begin{equation}
\beta(g)=a \frac{\partial g(a)}{\partial a}.
\label{380}\end{equation}
Equation (\ref{370}) is the RG equation.  As the present problem permits
analytic solution, the constant $\beta(g)$ of Eq. (\ref{380}) can be exactly
calculated.

For both ${\cal D}$ = 3, and 2, $\beta(g)$ is a finite quantity independent
of $a$.  For ${\cal D}=3$, from Eqs.  (\ref{x}), (\ref{355}) and (\ref{380})
we have $\beta(g) = -g - g^2.$ Similarly, for ${\cal D} =2$,  from Eqs.
(\ref{y}), (\ref{356}), and (\ref{380}) we have $\beta(g)=-g^2.$

One has the following Taylor series relating the  solution for  a small
finite $a$, and that for $a \to 0$:
\begin{eqnarray}
T(k,a)=T(k,0)+\frac {a^2}{2!}T''(k,0)+\frac{a^3}{3!}T'''(k,0)...,\nonumber
\end{eqnarray} 
where prime(s) denote derivative with respect to $a$ at $a=0$.  Here the
linear term in $a$ does not contribute, as the RG equation (\ref{360}) yields
$T'(k,0)=0$. Though the first order derivative is zero by the RG equation,
the higher-order derivatives are not zero.  Then the converged, $a\to 0$,
result is given, approximately, by
\begin{eqnarray}
T(k,0)\approx
T(k,a)-\frac {a^2}{2!}T''(k,a)-\frac{a^3}{3!}T'''(k,a)..., \label{uu}
\end{eqnarray} 
where the derivatives are to be calculated for  a small finite $a$. For
evaluating $T''$ ($T'''$) numerically one needs $T(k,a)$ for three (four)
adjescent values of $a$. As more terms  are maintained in Eq. (\ref{uu}) a
more converged $a \to 0$ limit is obtained.

In summary, we have used the lattice technique for solving the
nonrelativistic quantum scattering problem with delta potential in two and
three dimensions.  This technique leads to a lattice-regularized Green
function.  Finite physical result is obtained by employing standard
renormalization procedure with this regularized Green function as the
continuum limit is taken. The RG equation is written for this problem.
Lattice technique and RG equation should be valid for general nonrelativistic
potential models with ultraviolet divergence. Though we have illustrated the
lattice technique for scattering problems with ultraviolet divergences, it
should be applicable to any scattering problem. In fact, the present study
strongly suggests that, as in QED and QCD, with the use of modern computers
the lattice technique should be a powerful alternative tool for the
numerical solution of general nonrelativistic few- and many-body problems,
where, unlike in the present delta potential problems, analytic solutions can
not be formulated.

We thank  the Conselho Nacional de Desenvolvimento Cient\'{\i}fico e
Tecnol\'ogico, Funda\c c\~ao de Amparo \`a Pesquisa do Estado de S\~ao Paulo,
and Financiadora de Estudos e Projetos of Brazil for partial
financial support.

\end{document}